\documentclass[journal=nalefd,manuscript=letter]{achemso}

\usepackage{chemformula} 
\usepackage[T1]{fontenc}

\author{Ezra Alexander}
\date{\today}
\email{ezraa@mit.edu}
\author{Matthias Kick}
\author{Alexandra R. McIsaac}
\author{Troy Van Voorhis}
\affiliation{Department of Chemistry, Massachusetts Institute of Technology, Cambridge, Massachusetts 02139, USA}

\title{Understanding Trap States in InP and GaP Quantum Dots Through Density Functional Theory}


\begin{document}

\begin{tocentry}

\includegraphics[width=\textwidth]{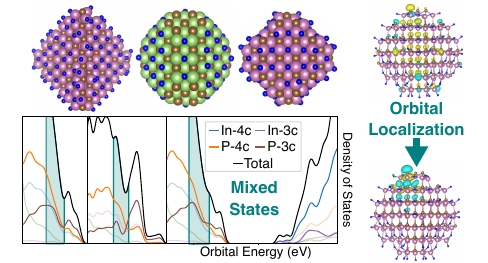}
For Table of Contents Only

\end{tocentry}

\begin{abstract}
The widespread application of III-V colloidal quantum dots (QDs) as non-toxic, highly tunable emitters is stymied by their high density of trap states. Here, we utilize density functional theory (DFT) to investigate trap state formation in a diverse set of realistically passivated core-only InP and GaP QDs. Through orbital localization techniques, we deconvolute the dense manifold of trap states to allow for detailed assignment of surface defects. We find that the three-coordinate species dominate trapping in III-V QDs and identify features in the geometry and charge environment of trap centers capable of deepening, or sometimes passivating, traps. Furthermore, we observe stark differences in surface reconstruction between InP and GaP, where the more labile InP reconstructs to passivate three-coordinate indium at the cost of distortion elsewhere. These results offer explanations for experimentally observed trapping behavior and suggest new avenues for controlling trap states in III-V QDs.
\end{abstract}

\section{}
Colloidal semiconductor nanocrystals, more commonly known as quantum dots (QDs), have attracted considerable attention as solution-processable materials \cite{dai_solution-processed_2014,garcia_de_arquer_solution-processed_2017} with highly tunable optical properties \cite{bawendi_quantum_1990,murray_synthesis_1993,alivisatos_semiconductor_1996,talapin_prospects_2010,boles_surface_2016,kagan_building_2016}. They have begun to see use in a wide range of applications including photovoltaics \cite{semonin_peak_2011,scalise_tailoring_2019}, photodetectors \cite{garcia_de_arquer_solution-processed_2017,livache_colloidal_2019}, LEDs \cite{dai_solution-processed_2014,won_highly_2019}, lasing \cite{fan_continuous-wave_2017,park_colloidal_2021}, drug delivery \cite{mura_stimuli-responsive_2013,patra_nano_2018}, biological imaging \cite{medintz_quantum_2005,saeboe_extending_2021}, and quantum computing \cite{ladd_quantum_2010,zajac_resonantly_2018}. However, the QDs with the best performance to date \cite{dai_quantum-dot_2017,hanifi_redefining_2019,garcia_de_arquer_semiconductor_2021} are those composed of highly toxic and internationally-restricted cadmium or lead chalcogenides \cite{derfus_probing_2004,reiss_synthesis_2016,allocca_integrated_2019}, making the development of a non-toxic alternative material with equally strong optical properties necessary for safe widespread commercialization. III-V QDs, namely indium phosphide (InP), are promising candidates for this replacement due to their low toxicity \cite{reiss_synthesis_2016,allocca_integrated_2019} and widely tunable emission range \cite{saeboe_extending_2021,reiss_coreshell_2009}. Until recently, their implementation has been held back by generally low quantum yields and broad emission line widths relative to their II-VI counterparts \cite{cros-gagneux_surface_2010,cui_direct_2013,tamang_chemistry_2016,hughes_effects_2019,jang_environmentally_2020}. These phenomena are often understood to result from a high density of trap states: occupied or virtual electronic states, usually localized on the surface of the QD, with energies between the valence band maximum (VBM) and conduction band minimum (CBM) \cite{hughes_effects_2019,giansante_surface_2017}. While recent advances in control over III-V core/shell heterostructures have led to InP QDs with near-unity quantum yields \cite{won_highly_2019,van_avermaet_full-spectrum_2022,li_znf2-assisted_2022}, a complete atomistic understanding of the formation and character of surface traps in III-V QDs remains elusive, especially for core-only QDs.

Trap states are not unique to III-V QDs; a vast body of both experimental and theoretical literature discusses trap states in II-VI QDs \cite{giansante_surface_2017,kirkwood_finding_2018,voznyy_dynamic_2013,houtepen_origin_2017,du_fosse_role_2019,goldzak_colloidal_2021,mcisaac_it_2023,kilina_effect_2009,wei_colloidal_2012,elward_effect_2013,califano_temperature_2005}, IV-VI QDs \cite{giansante_surface_2017,kim_impact_2013,chuang_open-circuit_2015,voros_hydrogen_2017}, and lead halide perovskite nanocrystals \cite{giansante_surface_2017,shao_origin_2014,nenon_design_2018,du_fosse_limits_2022}. The most widely accepted origin of trap states is under-coordinated surface atoms \cite{hughes_effects_2019,houtepen_origin_2017,nenon_design_2018,du_fosse_limits_2022,fu_inp_1997,stein_luminescent_2016,cho_optical_2018,kim_trap_2018,dumbgen_shape_2021,dumbgen_surface_2023,kilina_effect_2009}, although in certain systems excess charge \cite{voznyy_dynamic_2013,du_fosse_role_2019}, imperfect stoichiometry \cite{kim_impact_2013}, and substitutional defects \cite{janke_origin_2018} have also been implicated. For CdSe, several studies employing density functional theory (DFT) have shown that trap states arise primarily from two-coordinate Se atoms (Se-2c) but not Se-3c or any under-coordinated Cd \cite{houtepen_origin_2017,goldzak_colloidal_2021,mcisaac_it_2023}. No such consensus has been reached for InP QDs, however. Many studies have implied that hole trapping dominates in InP QDs \cite{kirkwood_finding_2018,stein_luminescent_2016,dumbgen_shape_2021,dumbgen_surface_2023,janke_origin_2018,calvin_thermodynamic_2020,richter_fast_2019,enright_role_2022}, but there is also considerable evidence for the presence of electron traps, especially in the absence of a core/shell heterostructure \cite{hughes_effects_2019,fu_inp_1997,cho_optical_2018,kim_trap_2018}. This disagreement has been compounded by a relative lack of atomistic \textit{ab initio} studies of trap states in InP QDs \cite{fu_inp_1997,cho_optical_2018,kim_trap_2018,dumbgen_shape_2021,dumbgen_surface_2023,janke_origin_2018,park_tuning_2021,ubbink_water-free_2022,hassan_electronic_2018}, many of which only employ less-accurate GGA functionals. Studies have variously emphasized P-3c traps \cite{dumbgen_surface_2023}, In-3c traps \cite{kim_trap_2018,hassan_electronic_2018}, both In-3c and P-3c traps but with disagreement on their respective depths \cite{fu_inp_1997,cho_optical_2018}, traps from the two-coordinate species with additional P-3c traps only in tetrahedral geometries \cite{dumbgen_shape_2021}, as well as studies that find InP QDs with both In-3c and P-3c to be trap free but see traps introduced upon different surface treatments \cite{janke_origin_2018,ubbink_water-free_2022}. Most of these studies only compute the electronic structure of a single model InP QD, limiting generalizability with respect to shape, size, faceting, and surface passivation. Moreover, very few studies have applied computation to understand trap states in other III-V QDs such as gallium phosphide (GaP) \cite{zhu_boosting_2023}, a promising but under-studied emissive material \cite{kim_origin_2014,choi_synthesis_2023}.

Here, we use DFT to study a  large, diverse set of InP and GaP QDs and draw generalized conclusions on the nature of their trap states and the factors that influence trap depth. The six base QD morphologies studied here are summarized in Figure 1. Several important decisions inform the development of our test set. We focus on core-only QDs, carved from the bulk crystal using a well-established construction procedure \cite{goldzak_colloidal_2021,geva_morphology_2018}.  We create six starting QDs for both InP and GaP, chosen to represent distinct faceting and  synthetically realizable shape\cite{kim_halideamine_2016,stein_probing_2018,kim_tailored_2021,zhao_engineering_2022,kim_shape-tuned_2022}. The four larger models, with diameters of 2-2.5 nm, represent the upper limit of computationally realizable QDs. The two smaller models allow for size extrapolation. Surfaces are passivated with X-type \ch{F-} ligands, representative of the well-established treatment of InP QDs with HF \cite{hughes_effects_2019,kim_trap_2018}. Calculations show that larger halogen ligands create states close to the VBM, potentially interfering with the assignment of trap and bulk states (SI I.I). As all our QDs are cation rich, one can equivalently think of them as stoichiometric \ch{InP}/\ch{GaP} cores passivated by Z-type \ch{InF3}/\ch{GaF3} ligands\cite{owen_coordination_2015}.

\begin{figure}[h]
\centering
\includegraphics[width=\textwidth]{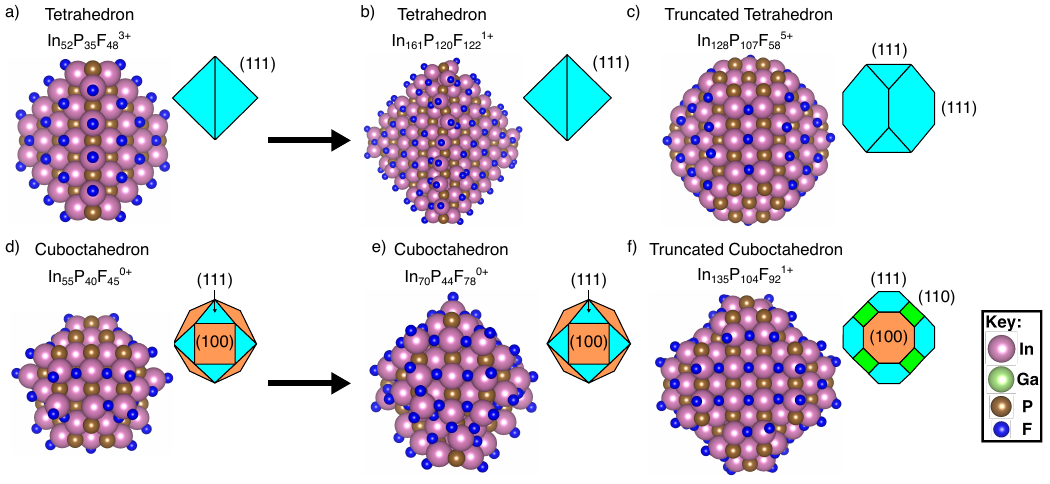}
\caption{Overview of the six base computational InP QD models used in this study. Each InP structure shown here has a GaP counterpart with the same shape and stoichiometry (SI, Figure S2). The colored inserts to the right of each structure provide a visual guide of the corresponding surface facets. Structures (b,c) and (e,f) can be thought of as extended versions of structures (a) and (d), respectively. }
\end{figure}

The construction procedure employed for these starting QDs is analogous to the ones used in previous \textit{ab initio} studies of trap states in III-V QDs (SI I.II), and results in some number of three-coordinate In/Ga and P atoms in all structures
\cite{kim_trap_2018,dumbgen_shape_2021,dumbgen_surface_2023,janke_origin_2018,ubbink_water-free_2022}. The difficulty in creating perfectly four-coordinate III-V model QDs arises from charge-orbital balance, in which the formal charge of each atom must add to the total charge of the system to prevent doping \cite{voznyy_charge-orbital_2012}. While some experimental evidence suggests QDs must be strictly charge-neutral \cite{fritzinger_utilizing_2010,moreels_size-tunable_2011}, it has also been shown that strict charge-balance greatly limits possible model III-V QDs, additionally restricting defects one could induce to these QDs to charge neutrality \cite{dumbgen_shape_2021}. To extend the range of QD shapes and defects available to us without inducing doping, we allow for slight positive charges in our structures \cite{voznyy_dynamic_2013,zherebetskyy_tolerance_2015}. The restriction to positive charges serves to avoid exacerbating DFT’s self-interaction error \cite{vydrov_tests_2007} and allows our cation-rich systems to have their charge balanced by fluoride counter-ions in solution \cite{stein_luminescent_2016}. We observe no doping in any of our systems and no qualitative difference in geometry or electronic structure between our charged and neutral models (SI I.III). 

We diversify our dataset and study the effects of surface reconstruction by creating "defective" QDs out of these starting models. These defective structures are created in a similar manner to previous studies, but our lack of charge neutrality affords us a greater variety of available defects \cite{fu_inp_1997,houtepen_origin_2017,giansante_surface_2017,zherebetskyy_tolerance_2015,dumbgen_shape_2021,cho_optical_2018,nenon_design_2018,mcisaac_it_2023}.
When creating defects in a starting structure, we consider all symmetry-unique removals of a single \ch{F-}, \ch{P^3-}, \ch{InF_x}, and \ch{InP}  (SI I.IV). This procedure results in a total library of 160 InP and GaP QDs for consideration. 

We then compute the ground state electronic structure of each QD using PBE0, as hybrid functionals are necessary for the accurate reproduction of band gaps \cite{kurth_molecular_1999,azpiroz_benchmark_2014}. Comprehensive identification of trap states requires the prior identification of the
 first bulk states, i.e. the VBM and CBM. This is made challenging by the dense manifold of intermediately localized states near the band edges (Figure 2). Two techniques are used to visualize the band structure of our models.  The first is the projected density of states (PDOS), which visualizes contributions from different atomic species to each band as a function of energy (Figure 2a,c). The second is the participation ratio (PR) (SI II.II), which measures the localization of each electronic state (Figure 2b,d). Analysis of the PDOS and PR alone is insufficient to understand the trap states in the QDs studied here for two reasons. First, the starting structures (Figures 1, S2) have trap states before any defects are induced, arising primarily from pre-existing three-coordinate atoms. Second, the trap states in our QDs are generally shallow because of their high degree of surface reconstruction. Thus, almost all QDs in our dataset display a dense quasi-continuum of trap states at both band edges, which obfuscate both new trap states induced by specific defects and the "true" CBM and VBM.

\begin{figure}[h]
\centering
\includegraphics[width=\textwidth]{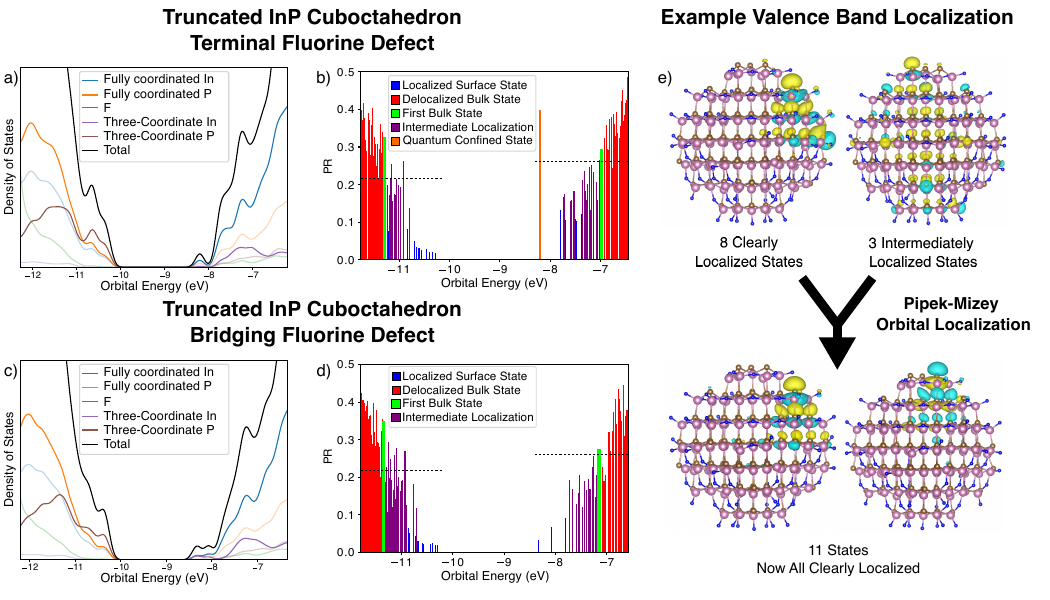}
\caption{(a,c) Projected density of states for two fluorine defects in the truncated InP cuboctahedron. Colored lines indicate contributions from different elements. (b,d) Participation ratio for the same two structures. Blue lines represent KS eigenstates that are clearly localized while purple lines represent intermediately delocalized states that become localized upon Pipek-Mizey localization. Green lines represent the first bulk state identified by our algorithm. The orange line represents the highly delocalized quantum-confined S-like state. (e) Pipek-Mizey orbital localization for the VB edge of the truncated InP cuboctahedron. Three states with intermediate localization are mixed with eight clearly localized states to form eleven localized states on different P-3c. Orbitals shown at an isosurface level of 0.03.}
\end{figure}

A naive analysis of the PR is problematic because localization is not an intrinsic property of DFT Kohn-Sham eigenstates \cite{boys_construction_1960,lehtola_unitary_2013}. In fact, any linear combination of degenerate eigenfunctions is also a solution to the Kohn-Sham equations, and in many systems these linear combinations will be more delocalized than what experiment and chemical intuition would suggest, especially when the eigenstate spectrum is particularly dense \cite{truhlar_are_2012}. Confronted with this problem, we can utilize orbital localization methods, such as Foster-Boys \cite{foster_canonical_1960} or Pipek-Mizey \cite{pipek_fast_1989}, to perform unitary transformations on a selected subset of molecular orbitals to maximize their localization. We find applying Pipek-Mizey localization to the band edges of our QDs reveals that many states with intermediate delocalization reduce to linear combinations of clearly localized trap states (Figure 2e). Combined with the observation that “true” bulk states fail to localize into clear surface states, orbital localization gives us a powerful tool to test for the location of the VBM and CBM. Our procedure is described in detail in the Supporting Information (II.III) alongside data highlighting the consistency of our predictions between related structures. As an example, the VBM is chosen to be the highest energy delocalized occupied state without a disproportionate contribution from under-coordinated P, and all occupied states above it in energy should be localizable into clearly surface-bound trap states. All such higher energy states are then taken to be hole traps, with trap depth equal to the difference between their energy and the energy of the VBM. An analogous definition identifies the CBM and associated electron traps.

A complication arises when considering the conduction band due to the intermittent presence of a low-energy, highly-delocalized state which likely corresponds to the $1S_e$ ground exciton state observed in experiment and the S-like envelope state predicted by "particle-in-a-sphere" theories \cite{bawendi_quantum_1990,norris_measurement_1996,ellingson_theoretical_2003,shulenberger_resolving_2021}. While delocalized,  this quantum-confined state cannot be considered the CBM due to its energetic isolation from the quasi-continuous conduction band. Furthermore, the state is present in less than 30\% of our structures. Figures 2b,d display two QDs with identical shape and stoichiometry but where the former displays the quantum-confined state, the latter does not. We have found no chemical justification for this intermittency. Nevertheless, our algorithm produces consistent VBMs, CBMs, and trap depths if we exclude the quantum confined state from consideration (SI II.IV). We observe no general qualitative difference between the trap states in QDs with and without the quantum-confined state. 

Recently, Snee \textit{et al.} found that DFT predicts non-singlet ground states for certain halide-passivated model InP QDs - a situation that would have major implications for the magnetic properties of these structures \cite{snee_dft_2021}. We were unable to demonstrate a similar effect in the structures studied here. For all 12 starting QDs in Figures 1 and S2, we find the singlet to be the lowest energy spin state, as would naively be expected for charge-balanced QDs. We thus conclude that there must be some important geometric differences between our QDs and those of Snee \textit{et al}. While it would be interesting to understand the nature of those differences, for now we focus on the non-magnetic ground states of these structures.

The defective QD structures reveal an immediate difference in surface reconstruction between InP and GaP (Figure 3a,b). While GaP is relatively rigid with little reconstruction after defect induction, InP reconstructs heavily both in the immediate vicinity of the defect and further afield. Representative of all QDs studied here, we use the smaller tetrahedral QD in Figure 3 as a particularly clear example of this effect. Degree of reconstruction can be quantitatively measured though reorganization energies, which we find to be over three times larger on average in InP than in GaP (SI III.I). We hypothesize the origin of this contrast to be In’s stronger, more ionic bonds with anionic ligands than Ga, due to In's lower electronegativity. We note that the observed trends in reconstruction still hold when Cl ligands are employed instead of F (SI III.II). These results are supported by prior experimental findings for InGaP, where Ga is found to reside disproportionately at the surface, and In-to-Ga substitution is found to be thermodynamically favorable, increasing QD stability and narrowing X-ray peaks \cite{lebedev_comparison_2016,hudson_synthesis_2022}.

\begin{figure}[!]
\centering
\includegraphics[width=\textwidth]{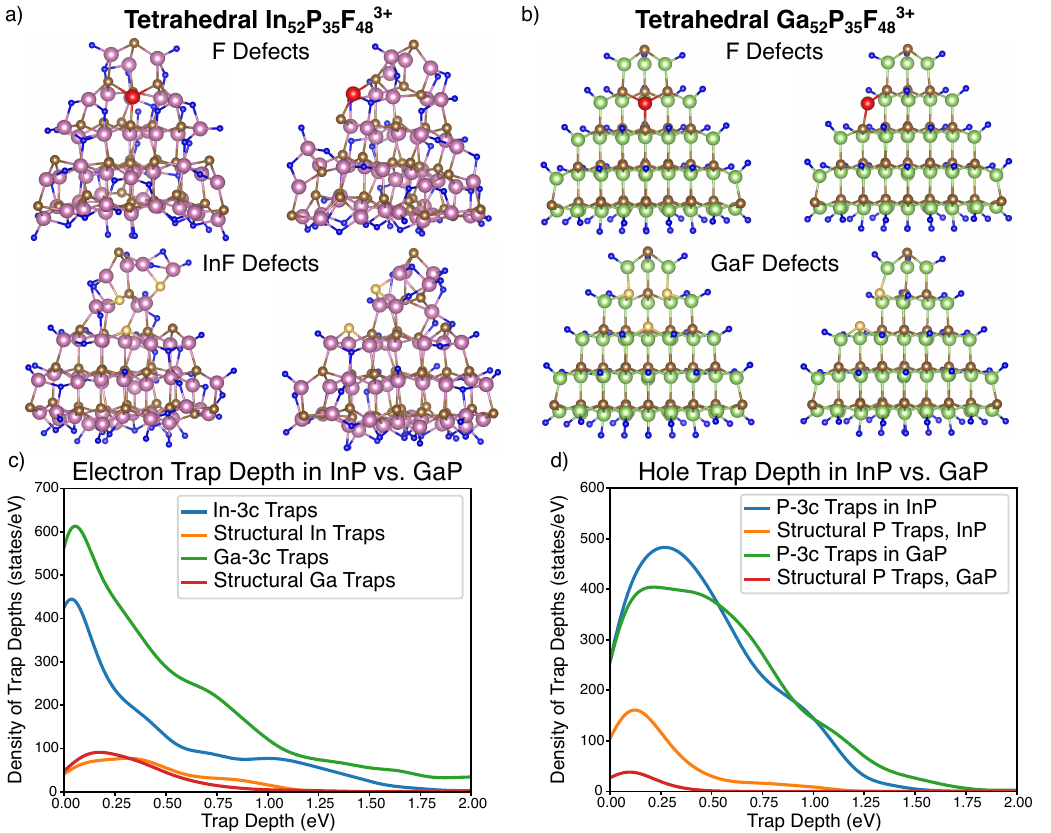}
\caption{ InP (a) and GaP (b) converged structures highlighting differences in surface reconstruction after corresponding F and InF defects are created in the smaller tetrahedra. Red colored atoms indicate the metal atom from which a fluorine anion is removed, and cream colored atoms indicate the phosphorus from which InF cations are removed. Distribution of trap depths for all electron (c) and hole (d) traps for all QDs in the dataset combined. Traps from InP and GaP, as well as 3-coordinate and structural traps, are colored separately. Discrete trap depths are broadened by normalized Gaussian functions with RMS width 0.1 eV. Non-trapping 3c atoms are assigned a depth of 0.}
\end{figure}

The greater reconstruction in InP leads to differences in the depth distribution of electron and hole traps between the two materials, shown in Figures 3c and 3d. Across our data set there are fewer In-3c in InP than there are Ga-3c in GaP, and similar numbers of P-3c in InP and GaP (SI III.I). The cost of the passivation of In-3c is evidenced in the formation of additional trap states localized around distorted In-4c and P-4c atoms, here denoted “structural” traps. While localized with mid-gap energies, these structural traps appear to arise not from under-coordinated surface species but rather from structural deformations caused by extensive surface reconstruction. We do not delve deeply into the nature and origin of these structural traps here; for now, we simply note their presence, even before orbital localization is applied. Despite these additional structural traps, electron traps in InP are generally less deep than those in GaP, with many In-3c being non-trapping or very shallowly trapping. This finding agrees with experimental results for InGaP nanowires, where nonradiative recombination increases with increasing Ga concentration \cite{zhang_carrier_2017}. P-based hole traps, on the other hand, have similar depths across InP and GaP QDs (Figure 3d). A structure with each defect clearly labeled can be found in the Supporting Information (III.III).

Close analysis of electron traps in InP and GaP QDs reveals two distinct geometries of the three-coordinate cation, which lead to distinct distributions of trap depths (Figure 4a,b). We designate these geometries as planar, when the cation is coplanar with its coordinated atoms, and pyramidal, when the cation is out of said plane. While both geometries are present in both InP and GaP, planar Ga-3c is far more prevalent relative to pyramidal Ga-3c than planar In-3c is to planar In-3c. This further explains the surface reconstruction in InP, where planar In-3c is more likely to convert to the pyramidal geometry  so ligands can bridge to additional In-3c. In both materials, traps arising from the pyramidal geometry are significantly deeper, on average, than those from the planar geometry. In InP this difference is such that most planar In-3c are effectively non-trapping. By separating the two geometries, we see that, averaging over all QDs in our dataset, planar Ga-3c (0.23 eV) forms deeper traps than planar In-3c (0.14
eV), and that pyramidal Ga-3c (0.84 eV) forms deeper traps than pyramidal In-3c (0.52 eV).

\begin{figure}[!]
\centering
\includegraphics[width=\textwidth]{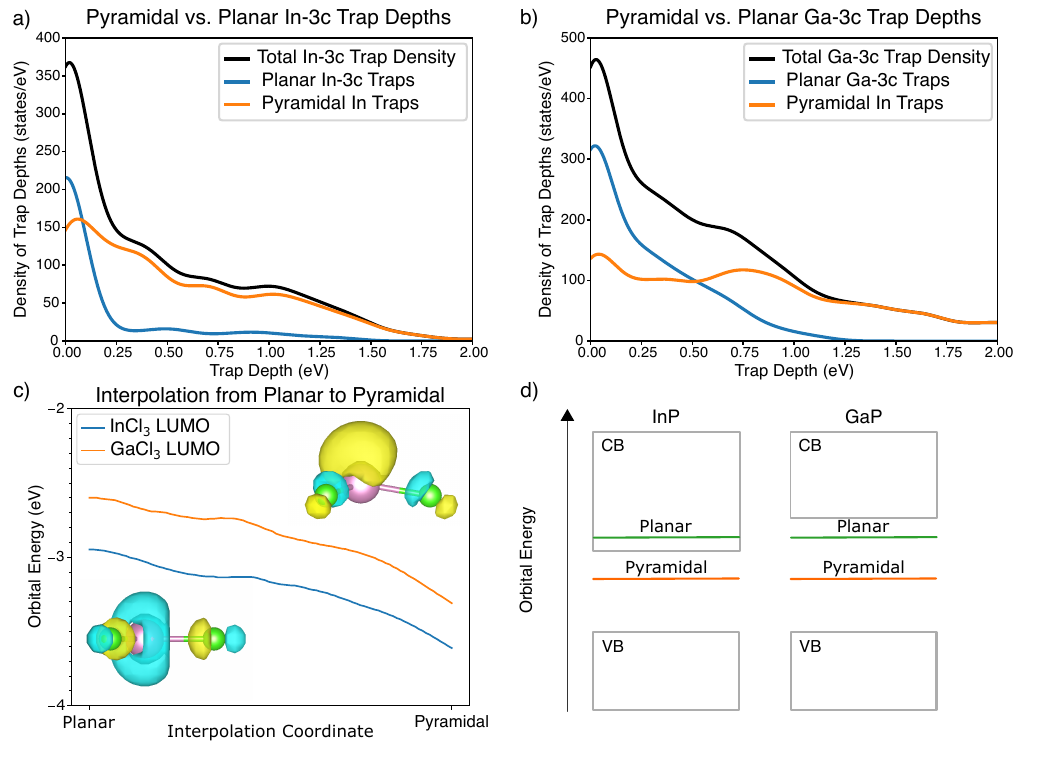}
\caption{Distribution of trap depths for In-3c (a) and Ga-3c (b) traps for all QDs in the dataset combined, with contributions from planar (blue) and pyramidal (orange) cations displayed separately. Without loss of generality, we only show the depth of the first trap on each 3c cation to better highlight trends. Discrete trap depths are broadened by normalized Gaussian functions with RMS width 0.1 eV. Non-trapping 3c atoms are assigned a depth of 0. (c) Change in LUMO energy when \ch{InCl3} and \ch{GaCl3} are interpolated from planar to pyramidal. Plots of initial and final \ch{InCl3} LUMO orbitals are inset with an isosurface level of 0.06. (d) Cartoon illustrating the difference in electron trap depths between InP and GaP arising from GaP's larger band gap.}
\end{figure}

To understand the origin of these differing trap depths, we performed interpolations between pyramidal and planar \ch{InCl3} and \ch{GaCl3} (Figure 4c). We find that the differences between pyramidal and planar defects can be understood using simple molecular arguments. As we interpolate from planar to pyramidal (SI IV.I), we observe that the LUMO, which corresponds to the trap state, decreases in energy by around 0.7 eV in both \ch{InCl3} and \ch{GaCl3}. This is accompanied by a shift of electron density onto the metal in the LUMO, as can be seen in the orbital plots in Figure 4c and through ChELPG charge analysis (SI IV.II) \cite{breneman_determining_1990}. This charge makes the LUMO more lone-pair-like, lowering the energy of the anti-bonding state resulting in a deeper trap. This analysis does not explain the difference between the electron trap depths in InP and  GaP, which most likely arise from GaP's wider band gap \cite{dumbgen_shape_2021,hudson_synthesis_2022} where shallow traps in GaP become non-trapping as the CBM decreases in energy (Figure 4d). We find support for this idea through additional interpolations between four-coordinate and three-coordinate \ch{InCl4} and \ch{GaCl4} (SI IV.III).

Even accounting for the differing depths of pyramidal and planar traps, the distributions of electron trap depths in Figure 4a,b remain quite broad. These shifts can be explained primarily through two electrostatic effects in the trap centers’ local environment. First, we observe that In-3c and Ga-3c bound to P-3c have shallower traps on average than those not bound to P-3c (Figure 5a,b). This can be understood by recognizing P-3c as having an excess of negative charge, which destabilizes the nearby trap state and raises its energy towards the CBM. A similar effect exists in reverse for hole traps, where P-3c bound to In-3c have shallower traps as the excess positive charge of the In-3c stabilizes the trap state (SI V). The second effect arises from the internal dipole moment of the QD. In both real QDs and our models, any asymmetry will lead to the creation of an internal dipole moment which can shift the depth of trap states, as previously noted for perovskite NCs \cite{du_fosse_limits_2022}. We show this effect for In-3c and Ga-3c in Figure 5c,d, computing the dipole overlap as the dot product between the dipole moment (pointing toward the positive charge) and the position vector of the trap center. We observe that a positive dipole overlap deepens trap states, whereas a negative dipole overlap almost always makes In-3c and Ga-3c nontrapping. Again, a similar effect exists in reverse for P-3c (SI V).

\begin{figure}[!]
\centering
\includegraphics[width=\textwidth]{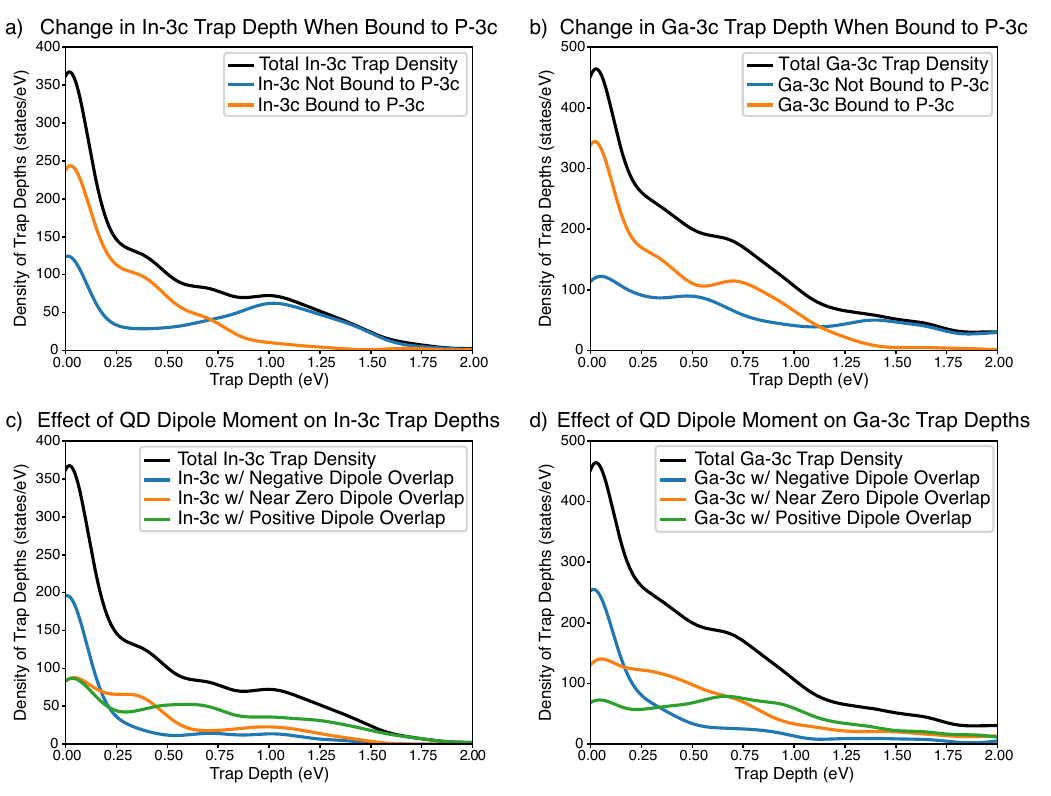}
\caption{(a,b) Trap depth distribution for In-3c and Ga-3c, respectively, across all QDs in the dataset combined with contributions from cations bound to P-3c separated from those not bound to P-3c. (c,d) Trap depth distribution for In-3c and Ga-3c, respectively, across all QDs in the dataset combined with contributions from cations with different dipole overlaps separated. Dipole overlap is calculated as the dot product of the position vector of the trap center with the dipole moment (pointing toward positive charge). A cutoff of (-40,40) is used to define the "near zero" region. Only the depth of the first trap on each 3c cation is shown  to highlight trends. Discrete trap  depths are broadened by normalized Gaussian functions with RMS width 0.1 eV. Non-trapping 3c atoms are assigned a depth of 0.}
\end{figure}

Note that we have not included traps from two-coordinate atoms in the above discussion. Such species appear in our dataset, and moreover cause deep traps when they appear (SI VI). However, we find that surface reconstruction in InP is sufficient to passivate most two-coordinate defects when created, and that structures with two-coordinate atoms are highly unstable relative to structures without. We thus conclude that two-coordinate species are at best minor contributors to trapping in real InP and GaP QDs.

In conclusion, we have investigated the prevalence, character, and depth of trap states in InP and GaP QDs using DFT. Our results are kept generalizable through the study of 160 QDs with variable size, shape, and surface defects. Through orbital localization, we deconvolute the dense band edge to identify otherwise evasive trap states including as yet unexplored “structural” traps tied to distorted fully-coordinated atoms. We leverage our dataset to analyze trends in trap depth arising from cation species, cation geometry, and local electrostatic effects which provide avenues for trap control. These results yield important insights into trap states in InP and GaP QDs, as well as informing guidelines for effective simulation of trap states in general QDs. Future directions include detailed investigation of the origin and character of the structural traps observed here, as well as the study of trap states in other III-V materials such as InGaP and III-V core-shell heterostructures. Investigation into the effects of surface oxidation and other impurities on trap state formation, as well as the efficacy of different surface passivation schemes, such as co-passivation with L-type ligands, is necessary to gain a complete picture of trap states in III-V QDs. Finally, it would be instructive to investigate the excited state electronic structure of III-V QDs directly using, for example, time-dependent DFT. 

\begin{acknowledgement}

This work was supported by the German Research Foundation (DFG, KI 2558/1-1). We would like to thank the MIT School of Science, MIT Chemistry Department, and the Office of Graduate Education for support of this work through the Dean of Science Fellowship.

\end{acknowledgement}

\begin{suppinfo}

The Supporting Information is available free of charge at XXX

\begin{itemize}
  \item Ligand choice, construction procedure, charge effects, defect induction procedure, computational details, participation ratio, procedure for identification of 1st bulk state, discussion of quantum-confined state, reconstruction energies, reconstruction with Cl ligands, labeled defects, interpolation details, ChELPG charge analysis, interpolation from four-coordinate to three-coordinate, broadening of hole trap depths, two-coordinate traps (PDF)
  \item All QD .xyz files (.zip)
  \item All trap depths (.xlsx)
\end{itemize}

\end{suppinfo}

\bibliography{bibliography}

\end{document}


\section{Structural Choices}

\subsection{Influence of Ligand on Band Structure}

To choose the ligands used in this study, we tested several halide ligands on our smaller cuboctahedral QD, shown in Figure S1. Geometries of each dot have been fully optimized with the PBE functional, and the band structure has been computed with the PBE0 functional \cite{adamo_toward_1999}. As one can see, fluorine contributed significantly less to the VB edge than chlorine and bromine. On the other hand, the chlorine and bromine passivated dots show a problematic density of ligand states in the energy region where the VBMs of the QDs are located, at -8.2 eV and -7.9 eV, respectively. Hence, to minimize potentially unphysical interference of these ligand states with the assignment of the first bulk state (SI II.III), we choose fluorine for use in the rest of this study. We note that fluorine sees extensive use in the passivation of InP QDs in experiment \cite{kim_trap_2018,hughes_effects_2019,won_highly_2019,li_znf2-assisted_2022,ubbink_water-free_2022}. We do not consider multi-atom organic ligands in this study, as their increased flexibility and electron count make geometry optimizations significantly more costly.

\begin{figure}[h]
\centering
\includegraphics[width=\textwidth]{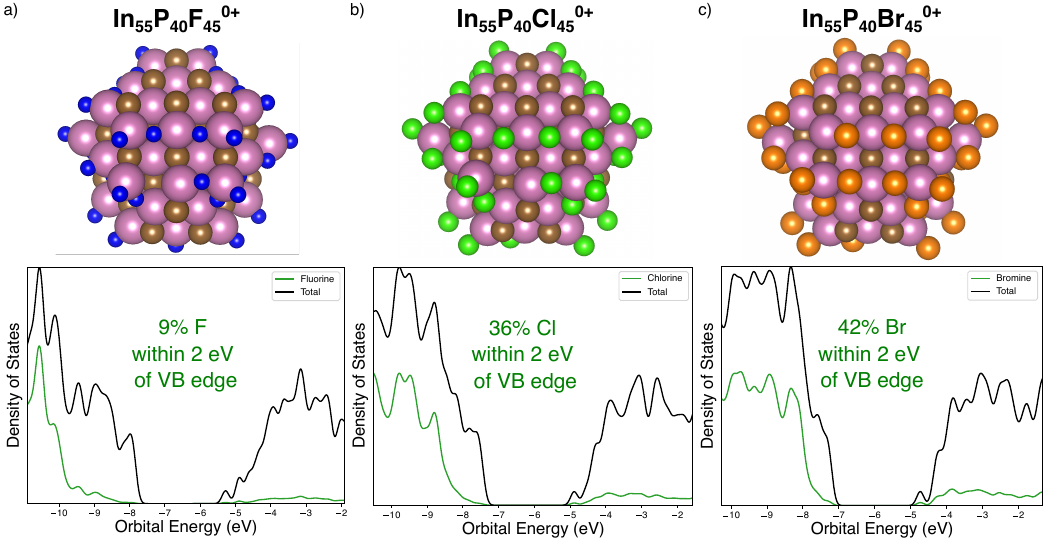}
\caption{Contribution of ligand states to the density of states for (a) fluorine, (b) chlorine, and (c) bromine. The structure of the dot is shown on the top, and the PDOS with all non-ligand lines omitted is shown on the bottom. Calculations are performed on the smaller cuboctahedral QD, individually optimized for each ligand. 2 eV is chosen as the cutoff for ligand impact as it corresponds to the deepest hole traps we see in our data set. Energies are given relative to vacuum.}
\end{figure}

\subsection{Construction Procedure}

To construct our QD models, we begin from the fully optimized InP and GaP bulk geometries. Choosing a point for the QD to be centered on (either P or In/Ga), we carve either a tetrahedral or quasi-spherical QD of a specified size from the bulk. We then prune any zero- and one-coordinate atoms from the QD. Any two-coordinate phosphorus are then turned into our passivating ligand, which we choose here to be fluorine (SI, Section I.I). We then add one additional ligand to each under-coordinated cation. This ensures that all species are at least three-coordinate. This procedure is inspired by and based on the well-established Wulff construction\cite{ringe_wulff_2011}, but differs in that we are not explicitly computing surface energies as our goal is not to find a single QD with optimal shape but rather create an ensemble of QDs of different shapes. In practice, we find that our procedure results in the majority of surface species being fully coordinated with some additional three-coordinate species of each type. Changing the center point and the initial size in our construction procedure allows us to generate QD models with different surface facets and cation-rich surface termination. All structures are then fully optimized until converged by the default CP2K geometry optimization thresholds at the PBE/def2-svp level of theory \cite{perdew_generalized_1996,metz_small-core_2000,weigend_balanced_2005}. The six InP structures created this way are shown in Figure 1, and their six GaP counterparts are shown in Figure S2.

\begin{figure}[h]
\centering
\includegraphics[width=\textwidth]{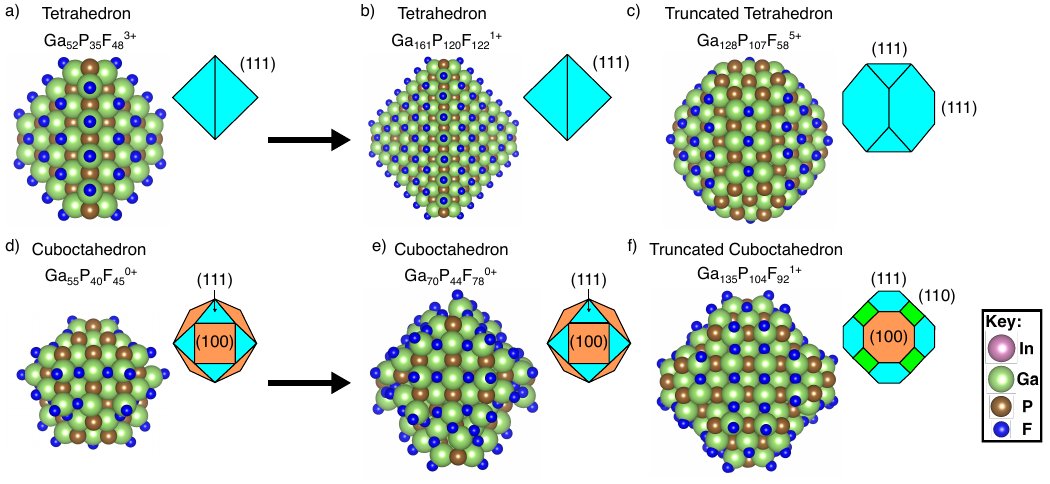}
\caption{Overview of the six base computational GaP QD models used in this study. The colored inserts to the right of each structure provide a visual guide of the corresponding surface facets. Structures (b,c) and (e,f) can be thought of as extended versions of structures (a) and (d), respectively.}
\end{figure}

\subsection{Charge Effects}

We observe no general difference between the band structure of the 30 charge neutral QDs in our dataset and the 130 positively charged QDs. That is because our positively charged QDs do not have excess positive charge, but rather the appropriate amount for the ions that form them as counted by the charge-orbital balance model \cite{voznyy_charge-orbital_2012}. To emphasize this point, we show in Figure S3 the band structure of two nearly identical QD structures with different charges. The structure on the left is formed by removing a +1 charged \ch{InF2} sub-unit from the +1 charged truncated cuboctahedral InP QD shown in Figure 1, forming two new P-3c and a QD that is net charge neutral. It has 18 hole traps, with a max hole trap depth of 0.91 eV, and 25 electron traps, with a max electron trap depth of 0.94 eV. The structure on the right is formed from the same base QD by removing instead an \ch{InPF} sub-unit, forming two new In-3c and a QD with a net charge of +2. It has 16 hole traps (two fewer than the other structure due to having two fewer P-3c) with a max trap depth of 0.70 eV, and 27 electron traps (again, from the two extra In-3c), with a max trap depth of 0.88 eV. The first bulk states in both dots appear at similar relative energies and are similar in localization and character. There is no general difference in the localization, depth, or number of trap states not easily explained by the QD structural differences. We note also that both structures display the quantum-confined state discussed further in section II.IV. Ten of the 30 charge neutral structures in our dataset display the state, roughly representative of the 37 of quantum-confined states in our 160 QD dataset.

\begin{figure}[h]
\centering
\includegraphics[width=\textwidth]{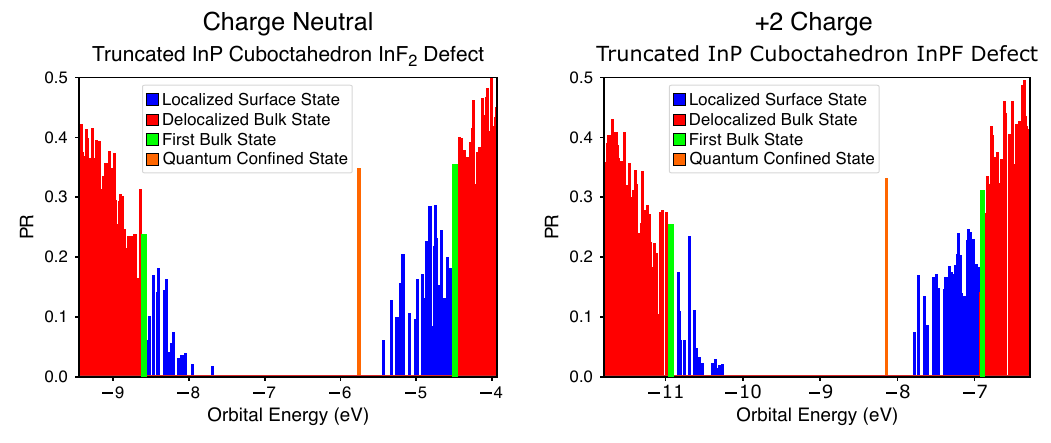}
\caption{Participation Ratio plots for two similar QDs, one that is charge neutral (left) and one that has a charge of +2 (right). The two QDs are different induced defects in the truncated cuboctahedral InP QD shown in Figure 1: an \ch{InF2} defect (left) and an InPF defect (right). The VBM and CBM found by our analysis procedure are bolded in green, trap states are blue, bulk states are red, and the quantum-confined S-like state is orange.}
\end{figure}

\subsection{Induced Defects}

We extend our data set from the 6 InP shown in Figure 1 and their 6 GaP counterparts shown in Figure S2 by inducing a number of different surface defects on our QDs, as exampled in Figure S4. Defects allow us to study the effects of surface reconstruction on our QDs and diversify the surface moieties present. Defects are created in all 12 starting QDs. Final defected dots are required to maintain a total charge that is neutral or slightly positive to avoid exacerbating DFT's self-interaction error \cite{vydrov_tests_2007}; for example, we do not create \ch{InF2} defects in the neutral smaller cuboctahedral QD as they would result in a QD with a -1 net charge.

For each charge-allowed type of defect listed in Figure S4, we then select each site that would create a distinct local configuration after the chosen atom or atoms are removed. The high symmetry of the QDs in our dataset make many such sites equivalent. For example, in the truncated cuboctahedron shown in Figure S4 we create three P-3c defects: two that create 3 In-3c and one that creates 2 In-3c and one In-2c. The two defects that create 3 In-3c are distinct in that one creates an In-3c bound to P-3c, and the other does not. A separate structure is then created where each selected defect is created, and then structures are fully optimized. This gives rise to our full test set of 160 QD structures. Through this creation of defects, our test set is tailored to give a generalized understanding of trap states beyond that accessible from a single QD structure, size, shape, faceting, or surface termination.

\begin{figure}[h]
\centering
\includegraphics[width=\textwidth]{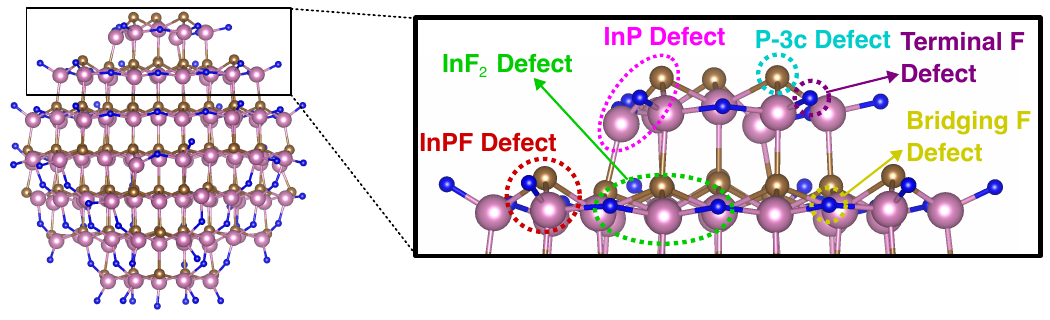}
\caption{Examples of each different type of defect induced in the truncated cuboctahedral InP QD. The name of the defect references the atoms removed in that defect. Due to the dot's high symmetry, the cutout represents all unique defect sites. All defects are created on the surface of the QD. Not shown here are  \ch{InF} and \ch{InF3} defects, which are made in some dots but not possible here.}
\end{figure}

\section{Computational Methodology}

\subsection{Computational Details}

All geometry optimizations were performed using the CP2K software package \cite{kuhne_cp2k_2020}. Optimizations employ the PBE functional and the DZVP-MOLOPT-GTH basis set. All QDs were optimized until converged by CP2K default thresholds. Preliminary test optimizations performed in Q-Chem result in near-identical structures to those generated by CP2K (though they take significantly longer to complete), and we identify near-identical ground states in Q-Chem for both the CP2K and Q-Chem optimized geometries.

All single-point band structure calculations were performed using the Q-Chem software package \cite{shao_advances_2015} as it enables the printing of molecular orbital Lowdin populations used in computing the PR (SI II.II). These calculations were performed with the PBE0 functional and def2-SVP basis set. A hybrid functional was chosen here as the incorporation of exact exchange is known to be necessary for the accurate reproduction of band gaps \cite{kurth_molecular_1999}. PBE0 in particular has been shown to perform quite well in benchmarking done on II-VI QDs \cite{azpiroz_benchmark_2014}. 

The double-zeta quality def2-SVP basis set is necessary to keep the computational cost of our calculations within the feasible regime. Preliminary test calculations performed on a subset of our smallest QDs with the smaller LANL2DZ basis set found it perform inadequately for these systems, due to its lack of polarizarion functions and treatment of more core electrons with the ECP. On the other hand, preliminary calculations conducted with the triple-zeta basis set def2-TZVP for a subset of our smallest QDs found no general qualitative differences with def2-SVP. Trap states undergo small shifts in energy of up to 0.1 eV, which causes changes primarily in the mixing of the dense manifold of trap states near the band edges. However, our orbital localization procedure yields the same overall set of localized trap states with both basis sets, and identifies the corresponding VBM and CBM. As such, we conclude the def2-SVP basis set to be sufficient for our needs. Def2-ECP effective core potentials are used throughout for indium.

\subsection{Participation Ratio}





The  participation ratio (PR) for a given QD eigenstate $i$ is computed according to the following formula:

\begin{align}
   {PR}_{\, i} \: = \: \frac{1}{N \: \sum_{j} \: q_{i,j}^{\,2}}
\end{align}

\noindent where $N$ is the total number of atoms and $q_{i,j}$ is the Lowdin population of atom $j$ in eigenstate $i$. This metric is bounded below by $\frac{1}{N}$ in the case where the state entirely arises from one atom and above by 1 in the case where the state is evenly spread over every atom. In practice, since valence band states arise primarily from phosphorus and conduction band states from indium/gallium, the most delocalized states we see have PR values of around 0.6.

We note here that many previous theoretical studies on trapping in QDs instead employ the inverse participation ratio (IPR), though the behavior and equation given for this quantity varies somewhat from work to work \cite{houtepen_origin_2017,goldzak_colloidal_2021,dumbgen_shape_2021,ubbink_water-free_2022,mcisaac_it_2023}. We make use of the PR instead here as it not only clearly differentiates localized states from delocalized states, but also better highlights the distinction between maximally delocalized states and intermediately delocalized states (as shown in Figure 2 in the main text). While highly localized states are clearly trap states, we find that some proportion of intermediately delocalized states in each QD are actually linear combinations of localized trap states. This can be revealed through the application of an orbital localization method, as described in the following section.

\subsection{Procedure for Identification of the 1st Bulk State}

Trap states are localized states with midgap energies. Identification of trap states thus requires the identification of the valence band maximum (VBM) and conduction band minimum (CBM). However, when the defect density of a QD is high, a high density of shallow trap states obscures the "true" bulk band edges in both the PDOS and the PR. This is because a high density of localized states can mix to form an equal number of unphysically delocalized states which are difficult to discern from true bulk states without the tedious process of observing the wavefunction of each state. We make use of information from the PDOS, from the PR, and from Pipek-Mizey orbital localization \cite{pipek_fast_1989}, chosen because of its preservation of $\sigma$-$\pi$ separation, to algorithmically identify the bulk band edges and thus all trap states in a given QD.

Our procedure for the identification of the VBM proceeds as follows. The first bulk state must satisfy two criteria: a low PDOS of under-coordinated atoms and a sufficiently high PR. For the PDOS, we require that the contribution from under-coordinated atoms be representative of their proportion in the QD:

\begin{align}
   D_{P-uc} (E) \: \leq \: c_{PDOS} \: \frac{N_{P-uc}}{N_{P-fc}} \: D_{P-fc} (E) \quad .
\end{align}

\noindent Here, $D_{P-u(f)c}(E)$ is the number of projected density of states of under (fully) coordinated phosphorus, $N_{P-u(f)c}$ is the number of under (fully) coordinated phosphorus in the QD, and $c_{PDOS}$ is a constant greater than 1. A broadening of 0.1 eV is used for the PDOS.

For the PR, we require that the first bulk state be delocalized above some proportion of the theoretical maximum PR of the valence band:

\begin{align}
   PR_{\, i} \: \geq \: c_{PR} \: \frac{N_{P}}{N} \: \frac{D_{tot} (E)}{D_{P} (E)} \quad .
\end{align}

\noindent Here, because the valence band is mostly composed of phosphorus states, the theoretical maximum PR of the valence band is taken as the proportion of atoms that are phosphorus ($\frac{N_{P}}{N}$) divided by the proportion of the total density of states that comes from phosphorus ($\frac{D_{P} (E)}{D_{tot} (E)}$). This corresponds roughly to a state delocalized over all phosphorus atoms plus a representative contribution from the other atomic species. $c_{PR}$ is a positive constant less than 1. The factors $c_{P DOS}$ and $c_{PR}$ allow for our algorithm to have some flexibility, and must be chosen empirically. While ideal values of these parameters in principle vary from structure to structure, we find that values of $c_{PDOS}=1.5$ and $c_{PR}=0.5$ reproduce expected results in all systems studied here.

Candidate bulk states which satisfy both criteria are vetted using orbital localization. The highest energy occupied state that satisfies both the PDOS and PR constraints is chosen as a candidate bulk state and screened through Pipek-Mizey orbital localization. If the candidate state is the true VBM, then all states above it in energy must be localizable trap states. If this localization yields any state which is not clearly localized on a single surface atom, then our thresholds have missed a higher energy bulk state. The thresholds are then lowered until a new, higher energy candidate is identified, and this process is repeated. When all states above the candidate in energy have been successfully localized, we must test that our candidate is in fact a bulk state and not just another mix of trap states. We do this by identifying the next state below our candidate in energy which satisfies both thresholds. An orbital localization is then applied to all states above this second candidate bulk state in energy. If our first candidate was a true bulk state, this localization should now fail to clearly localize every state, in which case our algorithm has successfully identified the VBM. If all states above this second candidate bulk state do successfully localize, then this state becomes our new primary candidate bulk state and the entire process is repeated. 

This approach successfully identifies a reasonable candidate bulk state for each system studied here. An analagous procedure can be performed to find the lowest energy bulk state in the conduction band by substituting P-3c for In-3c (or Ga-3c as appropriate) and moving upwards in energy instead of downwards. Table 1 summarizes the VBM and CBM energies predicted by our procedure. For a set of QDs with identical stoichiometry, for example the set of F defects in our larger cuboctahedral QD, one would expect the VBM and CBM energies to the be generally the same, with only the energies of surface traps associated with the specific defects fluctuating. Table 1 summarizes the average deviations of our predicted VBM and CBM energies from the other predictions for that QD shape and stoichiometry. The deviations in each case are quite small, below the accuracy threshold of our employed functional \cite{adamo_toward_1999}. We also highlight in Table 1 the effect of treating the quantum confined (QC) state, discussed in II.IV, as the CBM. Such an assignment renders our treatment of the conduction band highly unstable, especially in InP where the QC state is prevalent in around half the structures. This assignment would further eliminate 28.5\% of electron traps from our dataset. As the QC state is often the lowest energy unoccupied state, the depths of the remaining electron traps would not change significantly. However, analyses such as those performed in Figures 4 and 5 would be greatly complicated, as many of the 28.5\% of lost electron traps are nearly identical in terms of geometry and electrostatic environment to some of the remaining electron traps. In other words, nearly identical trap centers would sometime give rise to deep traps and other times have a depth of 0.

\begin{table}
\caption{Deviation in VBM and CBM Prediction}
\label{table:integral_values}
\begin{tabular}{l|cc|cc}
\hline
\hline
Band Edge & \multicolumn{2}{c}{Our Approach} & \multicolumn{2}{|c}{Using QC state as CBM} \\

  & MAD (eV) & RMSD (eV) & MAD (eV) & RMSD (eV) \\
\hline 
VBM (InP)        & 0.05   & 0.08 &-&-     \\
CBM (InP)      & 0.09 & 0.11 &0.22&0.28      \\
\hline
VBM (GaP)          & 0.05  & 0.06 &-&-         \\
CBM (GaP)      & 0.05 & 0.06  &0.13 &0.21       \\
\hline
\hline
\end{tabular}
\end{table}

\subsection{Quantum Confined State}

A greatly simplified but surprisingly accurate picture of the electronic structure of a quantum dot can be drawn by considering each carrier to be confined by a spherically symmetric infinite potential well, colloquially known as a "particle-in-a-sphere" model \cite{bawendi_quantum_1990,norris_measurement_1996,shulenberger_resolving_2021}. Such quantum-confinement arguments predict the LUMO of a QD to be delocalized with an S-like envelope function, with HOMO-LUMO gap decreasing with increasing dot size, both in agreement with experiment \cite{ellingson_theoretical_2003}. Experimentalists commonly denote this state, the electron component of the first excited state, $1S_e$. As the high computational cost of excited state methods like TD-DFT prevent their application to the systems studied here, our Kohn-Sham orbitals correspond better to the "particle-in-a-sphere" picture, and thus we refer to the state as the "quantum-confined state." 

Our analysis procedure ignores the quantum-confined state when it appears, as it does in only 37 out of the 160 structures considered here. Generally, the high-symmetry QDs shown in Figures 1 and S2 display the state, but QDs with induced defects do not, especially when there is significant surface reconstruction. We first identified the state in our calculations by visual inspection of plots of the Kohn-Sham LUMO, confirming its S-like symmetry. In our analysis procedure, we efficiently identify it simply when the LUMO, or a nearby unoccupied state, is highly delocalized. When it appears, we then exclude it both from any orbital localizations and candidacy as the CBM. Several factors inform this decision. By definition, the quantum-confined state is not a part of the quasi-continuous conduction band. Moreover, it would be technically quite difficult to use the quantum-confined state as our CBM as it is only sporadically present in our structures. Its energy is also highly dependent on QD size, making calculated trap depths for differently sized QDs incomparable. But even in a QD where the quantum-confined state is present, there are several mechanisms by which localized states above it in MO energy could act as trap states. Hot electrons relaxing to the quantum-confined state from the conduction band could and likely would get trapped along the way in these highly localized intermediate states. Furthermore, the localization of these states makes excited states involving them experience stronger electron-hole attraction, potentially lowering their excited state energies below that of the absorption onset. Further research is needed to explore these effects.

\section{Reconstruction in InP and GaP}

\subsection{Reconstruction Energies}

Qualitative data on the surface reconstruction in InP and GaP is presented in Table 2:

\begin{table}
\caption{Surface Reconstruction in InP and GaP QDs}
\label{table:integral_values}
\begin{tabular}{l|cccc}
\hline
\hline
Material  &    Av. \# In/Ga-3c  & Av. \# P-3c  &  Av. $\Delta E_{recon}$ (eV) \\
\hline
Pre-Defect InP      &   20.4  & 12.5  & -     \\
Post-Defect InP      &     12.2 & 13.3 & -4.72     \\
\hline
Pre-Defect GaP      &  23.1    & 13.2  & -         \\
Post-Defect GaP      &     19.4 & 13.6 & -1.44         \\
\hline
\hline
\end{tabular}
\end{table}

In the above, $\Delta E_{recon}$ is defined as:

\begin{align}
   \Delta E_{recon} \: = \: E_{relaxed} \: -  \: E_{unrelaxed} \quad .
\end{align}

\noindent Here, $E_{relaxed}$ is the total energy of the fully relaxed defected QD and $E_{unrelaxed}$ is the total energy of the QD after the defect has been induced but before the geometry has been re-optimized. Note that, even before defects are induced, there are more Ga-3c than In-3c and more P-3c in GaP than P-3c in InP. This is because, even though we carve identical starting structures from the bulk for the two materials, these structures are then fully optimized, and thus reconstruct differently. Note also that, although inducement of defects innately creates under-coordinated atoms, in both materials induced defects lead to surface reconstruction that passivates additional under-coordinated cations beyond those created by the defect. A significantly higher proportion of In-3c are passivated by this surface reconstruction in InP than Ga-3c in GaP, reflected by InP's larger average reconstruction energy. In general, the passivation of three-coordinate cations occurs through the transition of fluorine ligands from terminal to bridging binding modes. However, in some cases under-coordinated indium undergoes large displacement to reach full coordination at the expense of under-coordinating surface phosphorus, as evidenced in the increase in the average number of P-3c after surface reconstruction. These data imply that the increased surface reconstruction in InP is driven by the passivation of In-3c.

\subsection{Reconstruction with Cl Ligands}

A natural explanation for the difference in surface reconstruction between InP and GaP is indium's lower electronegativity than gallium, leading to a stronger affinity for electronegative ligands. Since we employ the highly electronegative fluorine in this study, one may then suspect that the difference between the two materials would be less pronounced with a ligand which binds less strongly. 

We test this question by creating alternate Cl-passivated versions of our InP and GaP smaller tetrahedral QD, the system which displays the most extreme surface reconstruction in InP. We then create an analagous set of defects in each of these dots, and measure their surface reconstruction against that seen in InP. The results of this investigation are summarized in Table 3 below.

\begin{table}
\caption{Surface Reconstruction with F ligands vs Cl ligands in Smaller Tetrahedral QD}
\label{table:integral_values}
\begin{tabular}{l|cccc}
\hline
\hline
Material  &    Av. \# In/Ga-3c  & Av. \# P-3c  &  Av. $\Delta E_{recon}$ (eV) \\
\hline
InP w/ F     &  4.1   & 7.4  &  -12.89     \\
GaP w/ F      & 22.1     & 5.1 & -1.40      \\
\hline
InP w/ Cl     & 3.6     & 5.6  &  -6.94        \\
GaP w/ Cl      &  14.6    & 5.1  & -2.59         \\
\hline
\hline
\end{tabular}
\end{table}

While the use of Cl ligands instead of F ligands clearly leads to several differences in surface reconstruction in these QDs, it remains clear that the surface reconstruction in InP is significantly stronger than that in GaP. Surface reconstruction in GaP QDs increases with the use of Cl ligands, with more Ga-3c being passivated and a higher reconstruction energy, but remains well below the degree of reconstruction in InP QDs. Interestingly, the switch to Cl ligands instead decreases the total reconstruction energy while decreasing the number of both three-coordinate species after surface reconstruction. Ligand affinity arguments alone would suggest that both materials should experience a decrease in surface reconstruction upon switch to a less electronegative ligand, which implies that the changes here arise primarily from the increased size of the Cl ligand affording easier bridging interactions. In fact, none of the Cl structures display the large amplitude motion present after reconstruction of the InP QDs with F, as reflected in their low number of P-3c. With Cl, InP is able to passivate more In-3c than it could with F without resorting to large amplitude motions, explaining the lower reconstruction energy. The smaller Ga-3c, on the other hand, can now more easily form bridging interactions, which explains the increase in GaP's reconstruction energy, although still not to the level of InP. 

\subsection{Labeled Defects}

Figure S5 displays a ball-and-stick model of one of the InP QDs in our datatset in which all of the prevalent different types trap-forming defects on one face are clearly labeled and color-coded. Cutouts of one of each of the main categories of trap-forming defects are provided. This structure (\ch{In160P119F121}\textsuperscript{2+}, created from the larger tetrahedral InP QD by inducing a \ch{InPF} defect), is chosen because only because it contains all five types of trap-forming defects in the same relative vicinity. Each type of defect is found in at least one QD of each base shape. Note that the cutout examples of each type of trap-forming defect are only generally representative of all defects of that type.

\begin{figure}[!]
\centering
\includegraphics[width=\textwidth]{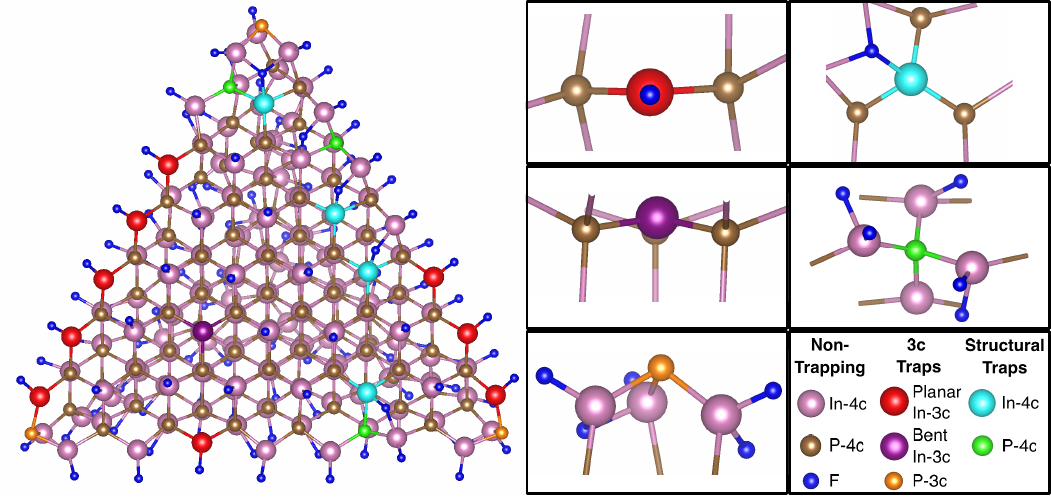}
\caption{Ball-and-stick model of a tetrahedral InP QD, in which all of the main categories of trap-forming defect on a single face of the nanocrystal have been clearly color-coded. Zoomed in cutouts of an example of each category of trap-forming defect have been provided on the right.}
\end{figure}

\section{Interpolations}

\subsection{Interpolation Details}

To explore the origin of the different trapping behavior between planar and pyramidal three-coordinate indium and gallium, we have investigated the electronic structure of a number of small, simplified In and Ga systems as they are interpolated non-linearly between different geometries. Idealized, fully symmetric versions of each endpoint are created and 98 intermediate structures are generated through a geodesic interpolation to ensure they remain within feasible space \cite{zhu_geodesic_2019}.  Two types of interpolation are performed for each material: one of \ch{InCl3}/\ch{GaCl3} from planar to pyramidal, and one of \ch{InCl4-}/\ch{GaCl4-} from tetrahedral to three-coordinate planar. In the latter, the four-coordinate to three-coordinate interpolation is accomplished by incrementally displacing the fourth Cl from the metal center until the total energy converges. Indium and gallium are coordinated with chlorine in these interpolations as its electronegativity is closer to phosphorus than that of fluorine.

All calculations on these simplified subsystems are carried out in Q-Chem \cite{shao_advances_2015}. DFT calculations on each structure are carried out with the PBE0 functional. To ensure even treatment of indium and gallium, the all-electron jorge-DZP-DKH basis set is employed \cite{jorge_contracted_2009,barros_gaussian_2010}. The entire interpolation trajectory is run in serial, with the orbitals from one frame being used as the initial SCF guess for the following frame to ensure smooth convergence. In the case of the \ch{InCl4-}/\ch{GaCl4-} interpolation from tetrahedral to three-coordinate planar, unrestricted calculations are employed to capture the singlet ground state of the separated planar geometry with a lone chlorine. The total energy varies smoothly from step to step for all interpolation trajectories. In each interpolation the lowest energy molecular orbital (LUMO) is taken to represent the electron trap that would be formed by the three-coordinate cation.

\subsection{ChElPG Charge Analysis}

To explain the energy trends observed in our interpolations from planar to pyramidal we utilize partial charge analysis, specifically charges from the electrostatic potential on a grid (ChElPG) to avoid strong basis set dependence \cite{breneman_determining_1990}. Again, these calculations are carried out in Q-Chem \cite{shao_advances_2015}. Since the LUMO "trap state" does not contribute to the ground state atomic charges, we approximate it's population as the difference in the atomic partial charge between the ground state and the double anion. Results are shown in Figure S6.

\begin{figure}[h]
\centering
\includegraphics[width=\textwidth]{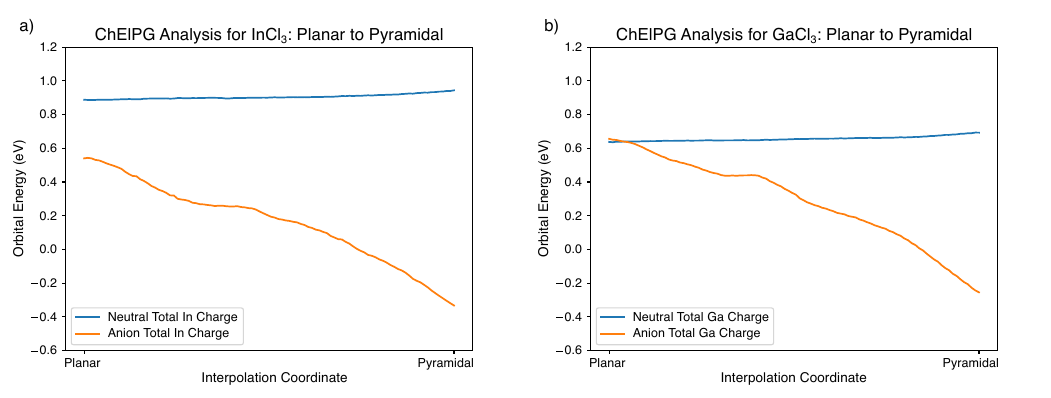}
\caption{ChElPG partial charge on (a) In-3c and (b) Ga-3c across interpolation from planar to pyramidal. The blue line represents the total charge on the metal in the neutral system, and the orange line represents the total charge on the metal in the -2 anion. The difference between the two lines serves as an estimate of the population on the metal in the LUMO trap state.}
\end{figure}

 Notably, in both materials the transition from planar to pyramidal leads to a marked increase in the population of the metal in the LUMO. Because the LUMO is anti-bonding in nature, the increase of the metal character makes the state more lone-pair like and thus decreases its energy, causing traps from pyramidal In-3c and Ga-3c to be deeper than traps from planar In-3c and Ga-3c. Note also that, regardless of geometry, these partial charges reproduce the expected difference in covalency between the two systems, with the more ionic \ch{InCl3} having both a more positive charge on indium in the ground state and more density on indium in the LUMO than the more covalent \ch{GaCl3}.

\subsection{Interpolation from Four-Coordinate to Three-Coordinate}

The aforementioned planar to pyramidal interpolations do not explain the difference in trap depth between In-3c and Ga-3c. To reference these MO energies to a common “bulk state”, we interpolate from a four-coordinate tetrahedron to the three-coordinate planar structure (Figure S7). We do this by incrementally displacing one of the four chlorine ligands while relaxing the rest of the structure to planar, until subsequent displacement of the lone \ch{Cl-} no longer significantly changes the total energy. Note that the lone chlorine necessitates these calculations be unrestricted, and starting at the 54th frame for InP and the 43rd frame for GaP the alpha and beta LUMOs have different energy.

In both cases the LUMO energy decreases drastically over the course of the interpolation, reflecting the lowering of the metal LUMO into the band gap to form a trap state. Moreover, we note that the decrease in both the alpha and Beta In LUMO energy is around 0.6 eV less than the decrease in the corresponding Ga LUMO energy. This agrees with the findings of our dataset, where electron traps in GaP are deeper than those in InP. This effect likely arises from GaP’s wider band gap, as discussed in the main text in Figure 4d. These interpolation energies qualitatively reproduce the ordering of average trap depths in our data set. Combining the two sets of interpolations, we qualitatively reproduce the ordering of electron trap depths in our dataset.

\begin{figure}[h]
\centering
\includegraphics[width=\textwidth]{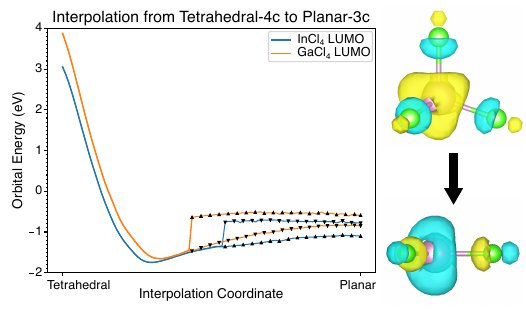}
\caption{Change in LUMO energy when \ch{InCl4} and \ch{GaCl4} are interpolated from a four-coordinate tetrahedral geometry to a three-coordinate planar geometry. Upwards and downwards facing triangles represent the spin +1/2 and spin -1/2 LUMO at geometries when they differ in energy. Plots of initial (top) and final (bottom) \ch{InCl4} LUMO orbitals are shown on the right with an isosurface level of 0.06.}
\end{figure}

\section{Broadening of Hole Trap Depths}

\begin{figure}[!]
\centering
\includegraphics[width=0.5\textwidth]{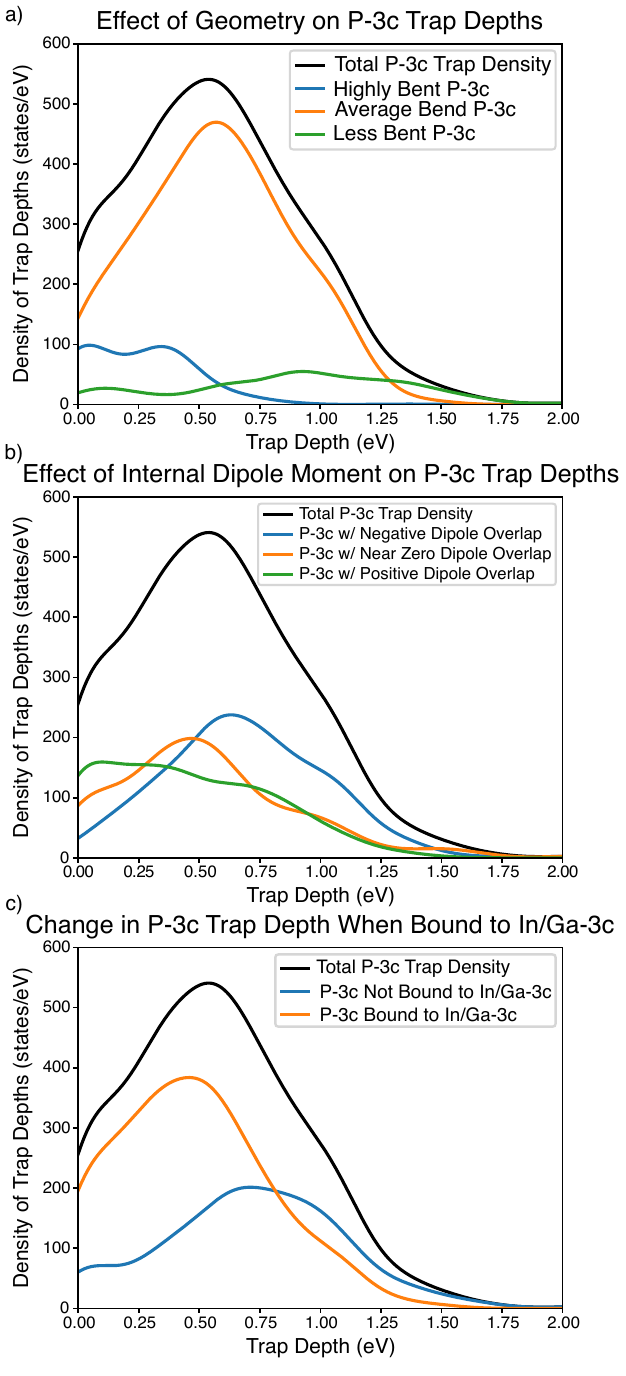}
\caption{Trap depth distributions for P-3c across all QDs in dataset with InP and GaP combined, with the shifts arising from different effects highlighted. Discrete trap depths are broadened by normalized Gaussian functions with RMS width 0.1 eV. (a) The effect of P-3c geometry. Average bend is defined as being within one standard deviation of the mean bend angle. (b) The effect of internal dipole moment. Dipole overlap is calculated as the dot product of the position vector of the trap center with the dipole moment (pointing toward positive charge). A cutoff of (-40, 40) $\si{\angstrom} \cdot$D is used to define the "near zero" region. (c) The effect of being bound to In-3c or Ga-3c.}
\end{figure}

Trends in the depth relative to the VBM of hole traps arising from P-3c are shown in Figure S8. Unlike the observed results for In-3c and Ga-3c, there are no distinct alternate geometries for P-3c in InP or GaP. All P-3c in our data set adopt a pyramidal geometry, with an average bend angle of $98.0^{\circ}$ and a standard deviation of $4.0^{\circ}$. Despite this relative uniformity, some trend does exist in hole trap depth when varying P-3c bend angle (Figure S8a). In general, more bent P-3c have shallower traps, and more planar P-3c have more deeper traps. This aligns with simple MO arguments - the $sp^2$ hybridized geometry has a non-bonding p-orbital which remains in the middle of the band gap. The shift of trap depth with internal dipole moment overlap is shown in Figure S8b. This trend operates in reverse from the trend observed for electron traps; here, negative charge deepens hole traps by destabilizing the electronic state to higher energies while positive charge stabilizes the hole trap, bringing its energy down to that of the VBM. The effect of being bound to In-3c and Ga-3c is shown in Figures S8c. Again, here we see the expected trend where In-3c serves as a local source of positive charge that stabilizes hole traps, reducing their depth.

\section{2-Coordinate Traps}

Two-coordinate atoms are the predominant source of trap states in CdSe QDs \cite{houtepen_origin_2017,goldzak_colloidal_2021,mcisaac_it_2023}, and some recent studies have indicated two-coordinate indium and phosphorus as sources of trap states in InP QDs \cite{dumbgen_shape_2021}. In-2c, P-2c, and Ga-2c are all formed in our QDs, and we find that these two-coordinate atoms create deep trap states when they appear. However, we do not focus on them in this work due to their high instability, making them unlikely to form in experiment. Data relating to the formation of two-coordinate trap states is summarized in Table 4 below. Results are for QDs with induced defects that, before the structure is allowed to relax, contain one or more two-coordinate species. $E_{unrelax}$ and $E_{relax}$ refer to the energy of the defect relative to the most stable isoelectronic defect in that QD before and after the structure is allowed to relax, respectively. $E_{unrelax}$ serves as a rough estimate of the barrier to defect formation, while $E_{relax}$, computed only for structures where a two-coordinate atom persists, measures the stability of such defects.

\begin{table}
\caption{Defects That Form Two-Coordinate Atoms}
\label{table:integral_values}
\begin{tabular}{l|cccc}
\hline
\hline
Species  &    \# Defects  & \# 2c after Reconst.  &  Av. $E_{unrelax}$ (eV) & Av. $E_{relax}$ (eV)  \\
\hline
In-2c     &  18    & 7  & 1.08 & 1.71     \\
P-2c in InP      & 18   & 3 & 1.25  & 2.93    \\
\hline
Ga-2c     & 18     & 16 &  1.27 & 1.82         \\
P-2c in GaP      & 18     & 9 & 1.08 & 2.38       \\
\hline
\hline
\end{tabular}
\end{table}

We see that surface reconstruction often passivates two-coordinate atoms in InP, but is less effective in GaP. However, defects that form two-coordinate atoms have a higher barrier to formation than corresponding defects which form three-coordinate atoms, and structures where two-coordinate atoms persist after surface relaxation are highly unstable. In general, this should prohibit the formation of two-coordinate atoms in both InP and GaP. A possible exception to this is Ga-2c, which is much more prevalent in our final dataset than other two-coordinate species. However, the relaxed and unrelaxed energies of Ga-2c containing QDs remain high enough relative to the most stable structures that we would still expect Ga-3c to be much more prevalent in GaP QDs than Ga-2c.

Note that, due to the relative instability of the defects that create 2c atoms, the proportion of structures in our dataset formed from such defects is if anything an overestimation of the prevalence of 2c surface species in real InP and GaP QDs. Note also that, unlike three-coordinate atoms, only two-coordinate Ga atoms form indirectly during surface reconstruction in our dataset.

\bibliography{bibliography}